\documentclass[aip,pra,twocolumn,showpacs]{revtex4}
\usepackage{graphicx,color}
\usepackage{amsmath}
\usepackage{longtable}
\begin{document}

\title{Simple estimation of thermodynamic properties of Yukawa systems}

\author{S. A. Khrapak,$^{1,2}$ A. G. Khrapak,$^2$ A. V. Ivlev,$^1$ and G. E. Morfill$^1$}
\date{\today}
\affiliation{ $^1$Max-Planck-Institut f\"ur extraterrestrische Physik, D-85741 Garching,
Germany\\ $^2$Joint Institute for High Temperatures RAS, 125412 Moscow, Russia}

\begin{abstract}
A simple analytical approach to estimate thermodynamic properties of model Yukawa systems is presented. The approach extends the traditional Debye-H\"{u}ckel theory into the regime of moderate coupling and is able to qualitatively reproduce thermodynamics of Yukawa systems up to the fluid-solid phase transition. The simplistic equation of state (pressure equation) is derived and applied to the hydrodynamic description of the longitudinal waves in Yukawa fluids. The relevance of this study to the topic of complex (dusty) plasmas is discussed.
\end{abstract}

\maketitle

\section{Introduction}

Yukawa systems are many-particle systems characterized by the pair interaction potential of the form
\begin{equation}\label{Yukawa}
U(r)= \epsilon (\sigma/r)\exp(-r/\sigma),
\end{equation}
where $\epsilon$ and $\sigma$ are the energy and length scales, and $r$ is the distance between two particles. This potential is often used to describe interactions in systems of charged particles immersed in a neutralizing medium. Two well known examples are colloidal dispersions and complex (dusty) plasmas~\cite{IvlevBook,FortovBook}. A remarkable property, explaining the significance of Yukawa systems in soft condensed matter research, is that by varying $\sigma$ it is possible to explore the extremely broad range of interaction steepness: From extremely soft Coulomb interactions ($\sigma\rightarrow \infty$) to very hard, almost hard sphere interactions ($\sigma\rightarrow 0$).

Various aspects of Yukawa systems have been the subject of study in the last several decades. This includes fluid-solid and solid-solid phase transitions and the emerging phase diagram, equilibrium transport properties, wave modes, confined systems and finite clusters, mixtures, various non-equilibrium phase transitions, etc. The attention has been focused on both two and three dimensional systems. It would be almost impossible to give credits to all related original works here, the reader is referred to books~\cite{IvlevBook,FortovBook} and some review papers~\cite{Lowen,FortovPR,MorfillRMP,Bonitz,Chaudhuri} instead.

Thermodynamic properties of Yukawa systems have also been extensively investigated. Molecular dynamics simulations~\cite{HamaguchiJCP1994,Farouki,HamaguchiJCP_1996,Hamaguchi} as well as integral equation theory (in the hypernetted chain approximation)~\cite{Kalman2000} have been used to calculate accurately the system energy. Since differentiations and integrations are required to obtain other thermodynamic quantities, their accurate determination remains a demanding computational task. Rather high accuracy is required in some cases. For instance, when locating fluid-solid and solid-solid phase transitions the free energies of respective phases have to be known with extreme accuracy since the smallest change in the free energy of either phase can result in a significant deviation from the actual coexistence line~\cite{Hamaguchi,DeWitt2001}. In some other cases no such accuracy is required and it would be valuable to have simple analytical expressions instead, which allow to estimate main thermodynamic properties of the system.

The purpose of this paper is to discuss such an approach for estimating thermodynamic properties of Yukawa systems. Although not extremely accurate, it is very simple and allows to catch the essential qualitative properties of these systems.

In the following we consider an idealized model consisting of point-like charged particles in the neutralizing surrounding medium. The mobile medium is responsible for screening so that the resulting pair interaction potential between the particles has the Yukawa form (\ref{Yukawa}). The main emphasize is on complex (dusty) plasmas and the relevance of the present idealized model to these systems will be discussed towards the end of the paper. In particular, we will point out that the model itself does not account for some important properties of complex plasmas. This implies that approximate analytical schemes which potentially can be extended to account for these properties are not irrelevant, although highly accurate data for an idealized model do exist.

The paper is organized as follows. In Section \ref{Model} we specify the model. In Section \ref{DH} we briefly remind the standard Debye-H\"{u}ckel approximation for weakly coupled Yukawa systems. The improvement of this model, which is the main subject of this paper, is described in Section \ref{DHH}. A limiting case of Yukawa systems -- the one-component plasma limit is briefly discussed in Section \ref{OCP}. We then proceed with the derivation of an approximate equation of state for Yukawa systems in Section \ref{EoS}. Its application to the analysis of waves in strongly coupled (fluid) Yukawa systems is described in Section \ref{Waves}. Section \ref{Concl} presents discussion and conclusion.

\section{Model}\label{Model}

We consider the two-component system consisting of microparticles of charge $Q$ and density $n$ and neutralizing
medium, characterized by the charge $-e$ and density $n_{\rm m}$ (subscript denotes ``medium''). In equilibrium the system is
quasineutral, so that
\begin{equation}\label{quasineutrality}
Qn_0 - en_{\rm{m}0}=0,
\end{equation}
where the subscript $0$ denotes unperturbed quantities. (If the neutralizing medium is comprised of several species and some
of them are oppositely charged, this should be taken into account: e.g., $n_{{\rm m}0}=n_{e0}-n_{i0}$ for an electron-ion
medium). It is conventional to characterize such a system by two dimensionless parameters:
\begin{equation}
\Gamma=\frac{Q^2}{a T} ~~~~{\rm and}~~~~ \kappa=ak_{\rm m},
\end{equation}
where $a= (3/4\pi n_0)^{1/3}$ is the Wigner-Seitz radius, $T$ is the system temperature (in energy units), and $k_{\rm
m}=\sqrt{4\pi e^2 n_{{\rm m }0}^{(\rm tot)}/T}$ is the inverse screening length (Debye radius) associated with the total
density of the neutralizing medium. In our case $n_{{\rm m}0}^{(\rm tot)}=n_{{\rm m}0}$, for the electron-ion medium $n_{{\rm m}0}^{(\rm tot)}=n_{i0}+n_{e0}$. In principle,
the particle species and surrounding medium can be characterized by different temperatures, but this is not important for
the present consideration. The coupling parameter $\Gamma$ is roughly the ratio of the Coulomb interaction energy, evaluated
at the mean interparticle separation, to the kinetic energy. The screening parameter $\kappa$ is the ratio of the
interparticle separation to the screening length.

Another inverse screening length-scale is associated with the particle component, $k_{\rm p}=\sqrt{4\pi Q^2 n_0/T}$. The
quantity $k_{\Sigma}=\sqrt{k_{\rm m}^2+k_{\rm p}^2}$ characterizes linear screening when both the particle and surrounding
medium are responsible for it. Note that $k_{\rm p}=\sqrt{3\Gamma}/a$ and, therefore, the relation between $k_{\Sigma}$ and
$k_{\rm m}$ takes the form $k_{\Sigma}=k_{\rm m}\sqrt{1+3\Gamma/\kappa^2}$.

The main quantities we will be dealing with in the following are the internal energy $U$, Helmholtz free energy $F$, and pressure $P$, associated with the particle component. In reduced units these are
\begin{equation}
u=U/NT,~~~ f=F/NT, ~~~ p=PV/NT,
\end{equation}
where $N$ is the number of particles in the volume $V$ (so that $n_0=N/V$).

\section{Debye-H\"{u}ckel approximation}\label{DH}

The Debye-H\"{u}ckel (DH) approximation corresponds to the limit of extremely weak coupling, $\Gamma \ll 1$. The electric field around a (test) particle is screened due to rearrangement of neutralizing medium and the particles themselves. Linearizing Boltzmann distributions for the both components and substituting this into the Poisson equation yields
\begin{equation}
\phi(r)=(Q/r)\exp(-k_{\Sigma}r)
\end{equation}
for the electrical potential distribution around the test particle. The reduced excess energy is
\begin{equation}
u_{\rm ex}=\frac{1}{2}\frac{Q}{T}\left[\phi(r)-\frac{Q}{r} \right]_{r\rightarrow 0}=-\frac{1}{2}\frac{Q^2k_{\Sigma}}{T}.
\end{equation}
This can be easily rewritten in terms of $\kappa$ and $\Gamma$ as
\begin{equation}
u_{\rm ex}(\kappa,\Gamma)=-\tfrac{1}{2}\Gamma\kappa\sqrt{1+3\Gamma/\kappa^2}.
\end{equation}
In the limit $\Gamma \rightarrow 0$ (where DH approximation is reliable) we have $u_{\rm ex}\simeq -\tfrac{1}{2}\Gamma\kappa$. In the one-component-plasma (OCP) limit, screening comes only from the particle component ($\kappa\rightarrow 0$) and we recover the familiar result $u_{\rm ex}\simeq -\frac{\sqrt{3}}{2}\Gamma^{3/2}$ ~\cite{Hansen}.

Other thermodynamic functions are easily obtained from the internal energy. For instance, the excess free energy in the fluid phase can be obtained via the integration
\begin{equation}\label{f}
f_{\rm ex}=\int_0^{\Gamma}d\Gamma'u_{\rm ex}(\kappa, \Gamma')/\Gamma'.
\end{equation}
In the DH approximation this integration is straightforward and yields
\begin{equation}\label{f_DH}
f_{\rm ex}(\kappa,\Gamma)=-\frac{\kappa^3}{9}\left[ \left(1+\frac{3\Gamma}{\kappa^2}\right)^{3/2}-1\right] .
\end{equation}
The reduced pressure is
\begin{equation}
p=n\left(\frac{\partial f}{\partial n}\right)_T.
\end{equation}
It is more convenient to rewrite this derivative in terms of $\kappa$ and $\Gamma$. In doing so we fix the density of the
surrounding medium and observe that
\begin{equation}
\frac{\partial \Gamma}{\partial n}= \frac{1}{3}\frac{\Gamma}{n} ~~~ {\rm and} ~~~\frac{\partial \kappa}{\partial n}= -\frac{1}{3}\frac{\kappa}{n},
\end{equation}
since $\Gamma\propto a^{-1}\propto n^{1/3}$ and $\kappa\propto a\propto n^{-1/3}$. The pressure then becomes
\begin{equation}\label{p}
p=\frac{\Gamma}{3}\frac{\partial f}{\partial \Gamma}-\frac{\kappa}{3}\frac{\partial f}{\partial \kappa}.
\end{equation}
Note that this definition is different from that used in Ref.~\cite{Farouki}, where the densities of neutralizing species (electrons and ions) were taken in constant proportion to $n$ (which resulted in the scaling $\kappa\propto n^{1/6}$). In the next section we will explain why the present choice seems more appropriate.

Applied to the excess free energy of the DH model this relation yields the excess pressure
\begin{equation}
p_{\rm ex}=-\frac{1}{2}\Gamma\kappa\left(1+\frac{3\Gamma}{\kappa^2}\right)^{1/2}+ \frac{\kappa^3}{9}\left[ \left(1+\frac{3\Gamma}{\kappa^2}\right)^{3/2}-1\right].
\end{equation}
In the OCP limit this reduces to the conventional expression $p_{\rm ex}= - \frac{1}{2\sqrt{3}}\Gamma^{3/2}=\frac{1}{3}u_{\rm ex}$.

The applicability of the DH approximation requires coupling to be small. To get an idea about its accuracy, let us compare the values of $f_{\rm ex}(\kappa, 1)$ (i.e. at $\Gamma=1$) calculated with the help of equation (\ref{f_DH}) with the ``exact'' numbers obtained from molecular dynamics (MD) simulations in Refs.~\cite{Hamaguchi,Farouki}. This comparison is shown in Table~ \ref{Tab1}. It is evident that even in the regime $\Gamma\lesssim 1$, the DH approximation is not characterized by high accuracy. Other approaches are needed and in the next Section we discuss one of the possible improvements.

\begin{table}
\caption{\label{Tab1} Reduced free energy in the weakly coupled regime, at $\Gamma=1$, for different values of $\kappa$. The second column corresponds to the ``exact'' values from MD simulations~\cite{Hamaguchi,Farouki}, the third column is computed using the Debye-H\"{u}ckel (DH) approximation (see Section \ref{DH}), and the last column is computed using the Debye-H\"{u}ckel plus hole (DHH) approach (see Section \ref{DHH}).}
\begin{ruledtabular}
\begin{tabular}{llll}
$\kappa$  & MD & DH  & DHH   \\ \hline
0.0 & -0.4368 & -0.577 & -0.460   \\
0.2 & -0.4495 & -0.588 & -0.471   \\
0.4 & -0.4809 & -0.617 & -0.502   \\
0.6 & -0.5284 & -0.660 & -0.548   \\
0.8 & -0.5866 & -0.715 & -0.606   \\
1.0 & -0.6541 & -0.778 & -0.673   \\
1.2 & -0.7304 & -0.848 & -0.747   \\
1.4 & -0.8103 & -0.922 & -0.826   \\
2.0 & -1.0710 & -1.169 & -1.084   \\
2.6 & -1.3504 & -1.435 & -1.360   \\
3.0 & -1.5424 & -1.619 & -1.549   \\
3.6 & -1.8326 & -1.900 & -1.838   \\
4.0 & -2.0274 & -2.091 & -2.033   \\
4.6 & -2.3223 & -2.380 & -2.326   \\
5.0 & -2.5200 & -2.574 & -2.523   \\
\end{tabular}
\end{ruledtabular}
\end{table}

\section{Debye-H\"{u}ckel plus hole approximation}\label{DHH}

The Debye-H\"{u}ckel plus hole (DHH) approximation allows to reduce inaccuracy of the Debye-H\"{u}ckel theory with respect to evaluating thermodynamics properties of moderately and strongly coupled OCP. The term ``Debye-H\"{u}ckel plus hole'' is conventionally associated with the work by Nordholm~\cite{Nordholm}, although similar arguments were used earlier~\cite{Iosilevskiy}.   The main idea behind the DHH approximation is that the exponential particle density must be truncated close to a test particle so as not to become negative upon linearization. Below we apply it to the model Yukawa system described in Section~\ref{Model}.

The main equations of this approximation are as follows. The electrical potential around a test particle is given by the
Poisson equation
\begin{equation}
\Delta \phi = - 4\pi(Qn -e n_{\rm m}).
\end{equation}
The neutralizing medium (each species when multi-component) follows the Boltzmann distribution, which can be linearized.
Other particles are absent in the sphere (hole) of radius $h$ around a test one. Outside the sphere, their density also
follows the Boltzmann distribution which can be linearized. This can be written as
\begin{equation}\label{dens}
n = \begin{cases} 0,  & r\leq h \\ n_0(1-Q\phi/T), & r>h. \end{cases}
\end{equation}
The quasineutrality condition (\ref{quasineutrality}) also holds.

For the potential inside the hole we then have $\Delta \phi_{\rm in}=k_{\rm m}^2\phi_{\rm in} + 4\pi en_{{\rm m}0}$ yielding
a general solution of the form
\begin{equation}\label{sol1}
\phi_{\rm in}(r)= ({\mathcal A}_1/r)\exp\left(-k_{\rm m}r\right) + ({\mathcal A}_2/r)\exp\left(k_{\rm m}r\right)+ {\mathcal A}_3,
\end{equation}
where ${\mathcal A}_3=-4\pi e n_{{\rm m}0}/k_{\rm m}^2=-3 Q/k_{\rm m}^2a^3$. Outside the sphere the potential satisfies
$\Delta\phi_{\rm out}= k_{\Sigma}^2\phi_{\rm out}$, which gives the following solution vanishing at $r\rightarrow\infty$:
\begin{equation}\label{sol2}
\phi_{\rm out}(r)= ({\mathcal B}/r)\exp\left(-k_{\Sigma}r\right).
\end{equation}
The two solutions, Eqs. (\ref{sol1}) and (\ref{sol2}), should be matched at the hole boundary, which yields $\phi_{\rm in}(h)=\phi_{\rm out}(h)$ and $\phi_{\rm in}^{\prime}(h)= \phi_{\rm out}^{\prime}(h)$. The two additional conditions are $\phi_{\rm out}(h)=T/Q$ (implying that the particle density vanishes at the hole boundary) and ${\mathcal A}_1+{\mathcal A}_2=Q$ (implying that $\phi_{\rm in}$ tends to $Q/r$ as $r\rightarrow 0$). This constitutes the full set of equations necessary to determine the hole radius $h$ as a function of $\kappa$ and $\Gamma$.

\begin{widetext}
Introducing the reduced hole radius $x=k_{\rm m}h$ we get after some algebra the following transcendent equation for
$x(\kappa,\Gamma)$:
\begin{equation}\label{hole}
x\kappa^2\left[\left(1+\sqrt{1+3\Gamma/\kappa^2}\right)e^{x}+\left(1-\sqrt{1+3\Gamma/\kappa^2}\right)e^{-x}\right]+ 3\Gamma\left[\left(x-1\right)e^{x}+\left(x +1\right)e^{-x}\right]-2\Gamma\kappa^3=0.
\end{equation}
The reduced excess energy is $u_{\rm ex}=\tfrac{1}{2}\tfrac{Q}{T}\left[({\mathcal A}_2-{\mathcal A}_1)k_{\rm b}+{\mathcal A}_3\right]$ which yields
\begin{equation}\label{u_ex}
u_{\rm ex}(\kappa, \Gamma)=\frac{x}{2}\left(1-\sqrt{1+3\Gamma/\kappa^2}\right)e^{-x}+\frac{3\Gamma}{2\kappa^2}\left(x+1\right)e^{-x}-\frac{3\Gamma}{2\kappa^2}-\frac{1}{2}\Gamma\kappa.
\end{equation}
\end{widetext}
The first two terms on the right-hand side of Eq.~(\ref{u_ex}) correspond to the particle-particle correlations in the DHH
approximation, the third term represents the excess free energy of the surrounding medium, and the last term is the free
energy of the sheath around each particle (see also Eq. (10) of Ref.~\cite{Hamaguchi}). Equation (\ref{u_ex}) can be
therefore rewritten as
\begin{displaymath}
u_{\rm ex}=u_{\rm pp}+u_{\rm m}+u_{\rm sh}.
\end{displaymath}

\begin{figure}
\includegraphics[width=8cm]{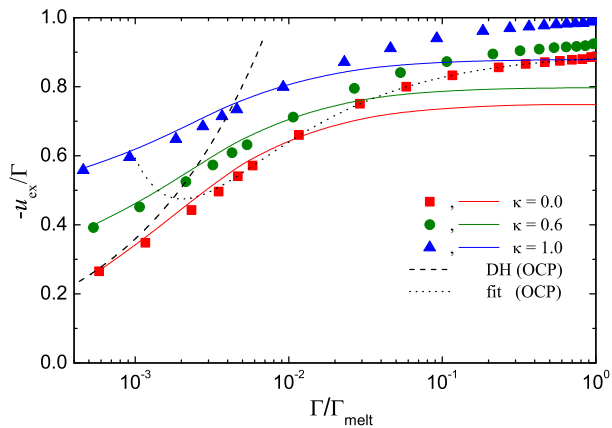}
\caption{(color online) Reduced excess energy (in units of $-\Gamma$) as a function of the reduced coupling parameter $\Gamma/\Gamma_{\rm melt}$ in the regime $\kappa\leq 1$. Solid curves correspond to our calculation using Eqs. (\ref{hole}) and (\ref{u_ex}), symbols are the results from MD simulations~\cite{Hamaguchi}. Data for $\kappa=0.0$, $0.6$, and $1.0$ are shown. The dashed curve corresponds to the DH approximation in the OCP ($\kappa=0$) limit. The dotted curve is the fit by Eq.~(\ref{fit}).}
\label{Uex1}
\end{figure}

Figure \ref{Uex1} shows the results of calculating the excess energy from Eqs. (\ref{hole}), (\ref{u_ex}) and comparison
with the numerical results obtained in Ref.~\cite{Hamaguchi} in the regime $\kappa\leq 1$. Here the reduced excess energy
$-u_{\rm ex}/\Gamma$ is plotted versus the reduced coupling parameter $\Gamma/\Gamma_{\rm melt}$, where $\Gamma_{\rm melt}$
is the coupling parameter at which fluid-solid phase transition occurs. The values of $\Gamma_{\rm melt}$ for a number of
$\kappa$ are tabulated in Table X of Ref.~\cite{Hamaguchi}; various analytical fits for the dependence $\Gamma_{\rm
melt}(\kappa)$ are also available~\cite{Hamaguchi,Vaulina,KM,JCP}. Figure \ref{Uex1} demonstrates that the DHH approximation
is rather accurate up to $\Gamma/\Gamma_{\rm melt}\sim 10^{-2}$. In this regime typical deviations of DHH from MD
simulations do not exceed few percent. For stronger coupling, DHH systematically overestimates the (negative) excess energy.
As fluid-solid phase transition is approached, the difference between DHH and MD simulations amounts to $\sim 15\%$ at
$\kappa=0.0$, and reduces to $\sim 10\%$ at $\kappa=1.0$.

\begin{figure}
\includegraphics[width=8cm]{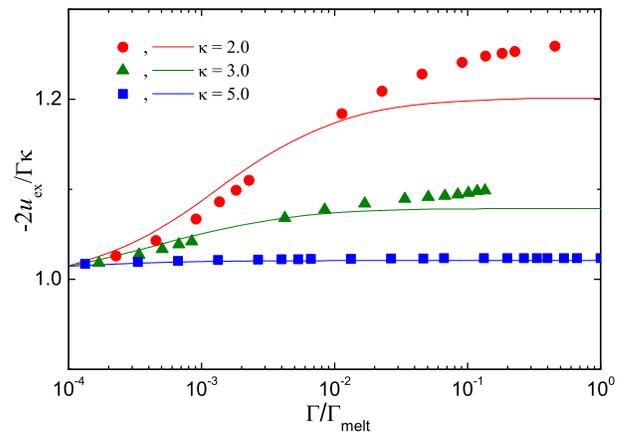}
\caption{(color online) Reduced excess energy (in units of $-\Gamma\kappa/2$) as a function of the reduced coupling parameter $\Gamma/\Gamma_{\rm melt}$ in the regime $\kappa > 1$. Curves correspond to our calculation using Eqs. (\ref{hole}) and (\ref{u_ex}), symbols are the results from MD simulations~\cite{Hamaguchi}. Data for $\kappa=2.0$, $3.0$, and $5.0$ are shown.}
\label{Uex2}
\end{figure}

Figure \ref{Uex2} shows the comparison between excess energies calculated using DHH approximation and obtained in MD
simulations in the regime $\kappa >1$. The qualitative picture remains the same as in the previously considered case. The
DHH approximation provides good accuracy up to $\Gamma/\Gamma_{\rm melt}\sim 10^{-2}$. Here the difference between DHH and
MD results does not normally exceed $\sim 1\%$. For stronger coupling DHH is again systematically overestimating $u_{\rm
ex}$. On approaching the fluid-solid transition the inaccuracy of DHH is $\sim 5\%$ for $\kappa=2.0$. As $\kappa$ increases,
the contribution from the sheath $u_{\rm sh}$ becomes dominant. In this regime $u_{\rm ex}\simeq u_{\rm
sh}\simeq- \tfrac{1}{2}\Gamma\kappa$ and DHH becomes virtually more and more accurate. Already for $\kappa=5.0$ one can hardly observe
any difference between DHH and MD simulations in Fig.~\ref{Uex2}.

The excess free energy can be calculated from Eqs. (\ref{u_ex}) and (\ref{f}).  To get an idea how DHH can improve the conventional DH approximation in the weakly coupled regime, the values of $f_{\rm ex}(\kappa, 1)$ have been calculated and listed in Table~\ref{Tab1}. Considerable improvement is evident. DHH approach underestimates the free energies $f_{\rm ex}(\kappa,1)$ from numerical simulations by approximately $5\%$ at $\kappa =0.0$ and $0.1\%$ at $\kappa=5.0$.

Overall, we observe that DHH approximation reduces to the DH theory in the limit of weak coupling ($\Gamma\ll 1$), but remains relatively accurate up to $\Gamma/\Gamma_{\rm
melt}\lesssim 10^{-2}$, where DH theory is already grossly wrong. For even stronger coupling the accuracy is merely qualitative. This is not surprising, since the approach
cannot catch the essential structural properties of the fluid state and thus cannot properly describe particle-particle
correlations. Consequently, the DHH approximation is completely useless when thermodynamic quantities have to be known with
sufficient accuracy (e.g. in the context of fluid-solid phase transition). In such cases direct numerical simulations are
required~\cite{Farouki,Hamaguchi}.  On the other hand, the appealing simplicity of the DHH approximation suggests to apply
it when no such accuracy is necessary. One particular example will be given in Section~\ref{Waves}.

The energy associated with particle-particle correlations can be also calculated via the energy equation~\cite{HansenBook}
\begin{equation}\label{energy}
u_{\rm pp}=\frac{2\pi n_0}{T}\int_0^{\infty}r^2V(r)g(r)dr,
\end{equation}
where $g(r)$ is the radial distribution function and $V(r)=(Q^2/r)\exp(-k_{\rm m}r)$ is the pair interaction energy. In the
limit of weak correlations the pair distribution function tends to unity everywhere, $g(r)=1$. In this case we easily get
$u_{\rm pp}=3\Gamma/2\kappa^2$ so that $u_{\rm pp}$ and $u_{\rm m}$ cancel each other exactly. The only remaining
contribution to the excess free energy is due to sheaths $f_{\rm sh}=u_{\rm sh}=-\tfrac{1}{2}\Gamma\kappa$. This will not give any
contribution to the pressure, as evident from Eq.~(\ref{p}) and thus $p_{\rm ex}=0$ if particle-particle correlations are
absent. This result, which has to be expected, critically depends on the model relation between $\kappa$ and $n$. In
particular, non-zero excess pressure would be obtained if the model of Ref.~\cite{Farouki} was used.

The DHH approximation is equivalent to the following form of $g(r)$ [cf. Eq.~(\ref{dens})]
\begin{equation}\label{g_ot_r}
g(r) = \begin{cases} 0,  & r\leq h \\ 1-Q\phi_{\rm out}(r)/T, & r>h. \end{cases}
\end{equation}
It is straightforward to verify that integration in Eq. (\ref{energy}) then yields the first two terms ($u_{\rm pp}$) in the right-hand side of Eq. (\ref{u_ex}).

%
%

\section{The OCP limit }\label{OCP}

The one-component-plasma is an idealized system of point charges immersed in a neutralizing uniform background of opposite charges. It corresponds to the limit $\kappa=0$
of the model under consideration. Various aspects of the OCP systems have been extensively studied in the literature. Among them are thermodynamic properties and, especially, the equation of state.     For the dependence of the excess energy on $\Gamma$ in the fluid phase it is conventional to use an expression of the form $u_{\rm ex}(\Gamma)=a\Gamma+b\Gamma^{s}+c+d\Gamma^{-s}$, which was obtained   using the variational hard sphere approach (yielding the exponent $s=1/4$)~\cite{DeWitt}. Later, it has been observed that the exponent $s=1/3$ yields somewhat better agreement with simulations~\cite{Stringfellow}. The resulting fit for the excess energy in the fluid phase as proposed in Ref.~\cite{HamaguchiJCP_1996} is
\begin{equation}\label{fit}
u_{\rm ex}(\Gamma)=-0.899\Gamma+0.565\Gamma^{1/3}-0.207-0.031\Gamma^{-1/3}.
\end{equation}
Figure~\ref{Uex1} demonstrates that it can be safely used in the regime $\Gamma/\Gamma_{\rm melt} \gtrsim 5\times 10^{-3}$,
i.e for $\Gamma\gtrsim 1$ (we remind that $\Gamma_{\rm melt}\simeq 170$ in the OCP limit). On the other hand the linear DH
approaxch is accurate only up to $\Gamma\simeq 0.01$ \cite{Hansen} and starts to overestimate significantly the actual
energy at $\Gamma\simeq 0.1$ (see Fig.~\ref{Uex1}).

Concerning the DHH approximation, the hole radius is directly obtained from Eq.~(\ref{hole}) expanding terms in series around $x=0$. This yields
\begin{equation}\label{hole1}
k_{\rm p}h=\left[1+(3\Gamma)^{3/2}\right]^{1/3}-1.
\end{equation}
The energy is then obtained from Eq.~(\ref{u_ex}), where two terms survive in the considered limit, $u_{\rm ex}=-\tfrac{1}{2}k_{\rm p}h-\tfrac{1}{4}k_{\rm p}^2h^2$. Using Eq. (\ref{hole1}) this becomes  \begin{equation}\label{ex1}
u_{\rm ex}=-\frac{1}{4}\left\{\left[1+(3\Gamma)^{3/2}\right]^{2/3}-1\right\}.
\end{equation}
Equations (\ref{hole1}) and (\ref{ex1}) coincide with those in Refs.~\cite{Nordholm,Iosilevskiy}. In the limit of very small $\Gamma$, Eq.~(\ref{ex1}) reduces to the DH result, but it remains adequate at much higher $\Gamma$ than the DH approach do. Figure~\ref{Uex1} demonstrates that the DHH approximation provides reasonable agreement up to $\Gamma\lesssim 1$, where the fit (\ref{fit}) starts to work. In the strongly coupled regime $\Gamma\gg 1$, the DHH approximation yields the correct scaling $u_{\rm ex}\propto \Gamma$, but the coefficient of proportionality is too low ($0.750$ instead of $0.899$). Note that in this strongly coupled regime the simplest ion-sphere model, yielding $u_{\rm ex}= -\tfrac{9}{10}\Gamma$~\cite{Baus}, reproduces the leading term of the fit (\ref{fit}) with impressive accuracy.

To conclude this section we point out that the hole radius defined by Eq.~(\ref{hole1}) represents the distance of the minimum separation between the OCP particles in the DHH approximation.
The particle-particle interaction can thus be viewed as the strong short-range hard-sphere repulsion at $r\leq h$ plus weak long-range Debye-H\"{u}ckel repulsion at $r>h$. It is reasonable to apply linear plasma response formalism to describe momentum transfer in distant collisions between OCP particles. In doing so the inverse hole radius $h^{-1}$ should be used as an estimate of the maximum wave vector $k_{\rm max}$ entering into the kinetic definition of the Coulomb logarithm. The resulting {\it effective} Coulomb logarithm reduces to the classical expression in the regime of weak coupling, but remains meaningful for strong coupling, too. It has been recently shown that such an approach would describe reasonably the relaxation rate and the self-diffusion coefficient of OCP over the entire region of coupling, up to the fluid-solid transition~\cite{CoulLog}.

\section{Towards an equation of state}\label{EoS}

In this section we derive an equation for the excess pressure of the particle component in the DHH approximation. As has
been discussed earlier, the excess free energy of the sheath does not contribute to the excess pressure. The
excess pressure arising from the particle-particle correlations can be conveniently evaluated from the virial pressure
equation~\cite{HansenBook}. Since the terms corresponding to particle-particle correlations and neutralizing medium cancel
each other exactly in the limit of no correlations [$g(r)=1$], the resulting expression for the excess pressure becomes
\begin{equation}
p_{\rm ex}=-\frac{2\pi n_0}{3T}\int_0^{\infty}r^3V^{\prime}(r)\left[g(r)-1\right]dr .
\end{equation}
Combining with the expression (\ref{g_ot_r}) for $g(r)$ in the DHH approximation we get after some algebra
\begin{multline}\label{p_ex}
p_{\rm ex}= \frac{1}{2}\frac{\Gamma}{\kappa^2}\left[-3 +e^{-x}\left(3+3x+x^2\right)-  \right. \\ \left. -\frac{e^{-x}}{1+\sqrt{1+3\Gamma/\kappa^2}}\left(x+x^2+\frac{x}{1+\sqrt{1+3\Gamma/\kappa^2}}\right)\right].
\end{multline}

\begin{figure}
\includegraphics[width=8cm]{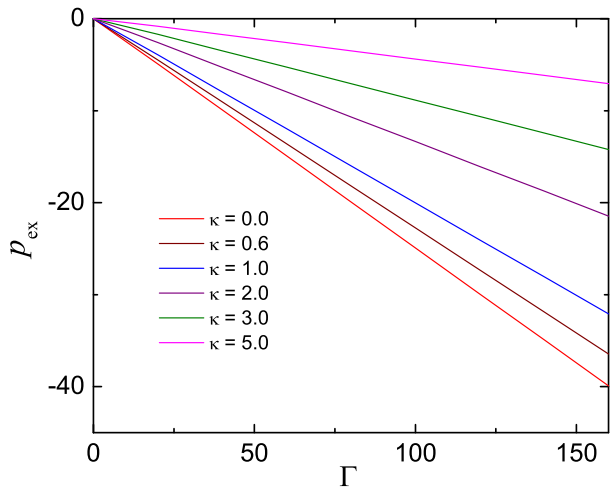}
\caption{(color online) Reduced excess pressure, $p_{\rm ex}$, as a function of the coupling parameter $\Gamma$ in the DHH approximation. Data for various $\kappa$ are shown. $p_{\rm ex}$ increases with $\kappa$.}
\label{Pex2}
\end{figure}

Figure \ref{Pex2} shows the calculated $p_{\rm ex}$ as a function of $\Gamma$ for several values of $\kappa$ (same values as
used in Figs.~\ref{Uex1} and \ref{Uex2}). Excess pressure is always negative and decreases as the fluid-solid transition is
approached. For $\kappa\lesssim 1$ the pressure curves are relatively close to each other, as has been already pointed out
in Ref.~\cite{Farouki}. This suggests to use a more accurate fit based on Eq.~(\ref{fit}) in this regime (not shown in the
Figure). With further increase in $\kappa$, the excess pressure tends to be less and less negative at a given value of
$\Gamma$. (Note however that comparison in terms of $\Gamma/\Gamma_{\rm melt}$ would apparently be more appropriate here).
Thus, the OCP fit (\ref{fit}) becomes completely irrelevant for $\kappa\gtrsim 1$. We will discuss this issue further in the
context of the particle density waves in Yukawa systems (dusty plasmas).

\section{Dust Acoustic Waves}\label{Waves}

The minimalistic model for the dust acoustic waves (DAW)~\cite{DAW} in dusty plasmas yields the following dispersion relation
\begin{equation}\label{DR}
\frac{\omega^2}{\omega_{\rm p}^2}=\frac{q^2}{q^2+\kappa^2}+\frac{q^2}{3\Gamma}\gamma\mu_{\rm p},
\end{equation}
where $\omega$ is the wave frequency, $\omega_{\rm p}=\sqrt{4\pi Q^2n_0/m_p}$ is the frequency-scale associated with the particle component (dust plasma frequency), $q=ka$ is the reduced wavenumber, $\gamma=C_p/C_v$ is the adiabatic index, and $\mu_{\rm p}=(1/T)(\partial P/\partial n)_{T}$ is the inverse reduced isothermal compressibility~\cite{Kaw1998}. The inverse compressibility $\mu_{\rm p}$ is related to the excess pressure via
\begin{equation}\label{compr}
\mu_{\rm p}= 1+p_{\rm ex}+\frac{\Gamma}{3}\frac{\partial p_{\rm ex}}{\partial \Gamma}-\frac{\kappa}{3}\frac{\partial p_{\rm ex}}{\partial \kappa},
\end{equation}
The dispersion relation (\ref{DR}) can be easily derived using the Boltzmann response of the neutralizing medium along with
the simplest hydrodynamic description of the particle component.  Substituting these into the Poisson equation and
linearizing will immediately yield Eq.~(\ref{DR}). In complex (dusty) plasmas the neutralizing medium normally consists of
positively charged ions and negatively charged electrons. Each component can be characterized by its own temperature, but
this is not essential for the present consideration, since only the actual value of $\kappa$ is affected. Note that in the
limit $\kappa=0$ Eq.~(\ref{DR}) coincides with the phenomenological hydrodynamic dispersion relation of the OCP model (see
e.g. Eq. (4.51) from Ref.~\cite{Baus}; in this case, the leading terms yield $\mu_{\rm p}\simeq 1+\frac49u_{\rm
ex}$). Note also that in the original derivation of Ref.~\cite{DAW} the particle component pressure was neglected at all,
corresponding to $\gamma\mu_{\rm p}/\Gamma=0$ (i. e. the assumption of cold particles was used).

Another dispersion relation suggested in the literature for DAWs in the strongly coupled regime reads~\cite{Kaw1998,Kaw2001}
\begin{equation}\label{OCP_disp}
\frac{\omega^2}{\omega_{\rm p}^2}=\frac{q^2}{q^2+\kappa^2}+\frac{q^2}{3\Gamma}\left(3+\tfrac{4}{15}u_{\rm ex}\right),
\end{equation}
This type of dispersion relation originates from the sum-rule analysis of the OCP in the long-wavelength limit (e.g. Eq.~(4.52) from Ref.~\cite{Baus}; see also Eq. (21) from Ref.~\cite{Abramo}).

\begin{figure}
\includegraphics[width=7cm]{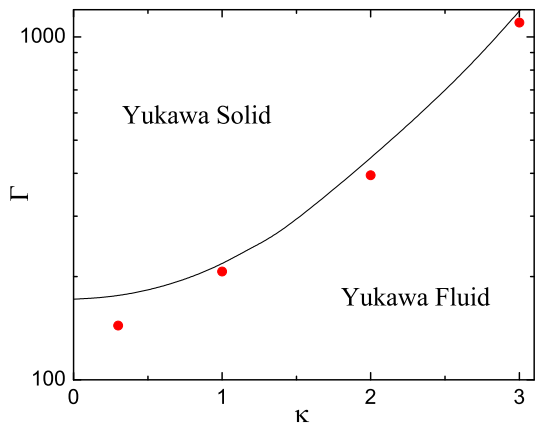}
\caption{(color online) Sketch of the phase diagram of Yukawa systems in ($\kappa$, $\Gamma$) plane. The curve marks the fluid-solid phase transition as obtained in Ref.~\cite{Hamaguchi}. The symbols correspond to the phase states for which the numerical simulations of density waves have been performed. The dispersion relations for the longitudinal waves at these state points are shown in Fig.~\ref{Dispers}.}
\label{PD}
\end{figure}

In the context of complex plasmas, both dispersion relations (\ref{DR}) and (\ref{OCP_disp}) neglect a number of properties specific to these systems, including e.g. collisions between different components (of particular importance are particle-neutral collisions), particle charge variations, external and internal forces acting on the particles (except the electrical one), etc.
They are however appropriate for comparison with idealized computer experiments designed to study the effect of strong coupling on wave dispersion in Yukawa systems. We take the results of Ref.~\cite{MD_DAW}, where wave dispersion relations in the fluid phase of Yukawa systems were obtained using molecular dynamics simulations. Simulations were performed for several state points characterized by certain values of $\Gamma$ and $\kappa$ parameters. These state points are shown in Fig.~\ref{PD}, representing the sketch of the phase diagram of Yukawa systems. All the investigated state points correspond to rather strong coupling -- they are located just below the melting curve.

\begin{figure*}
\includegraphics[width=15cm]{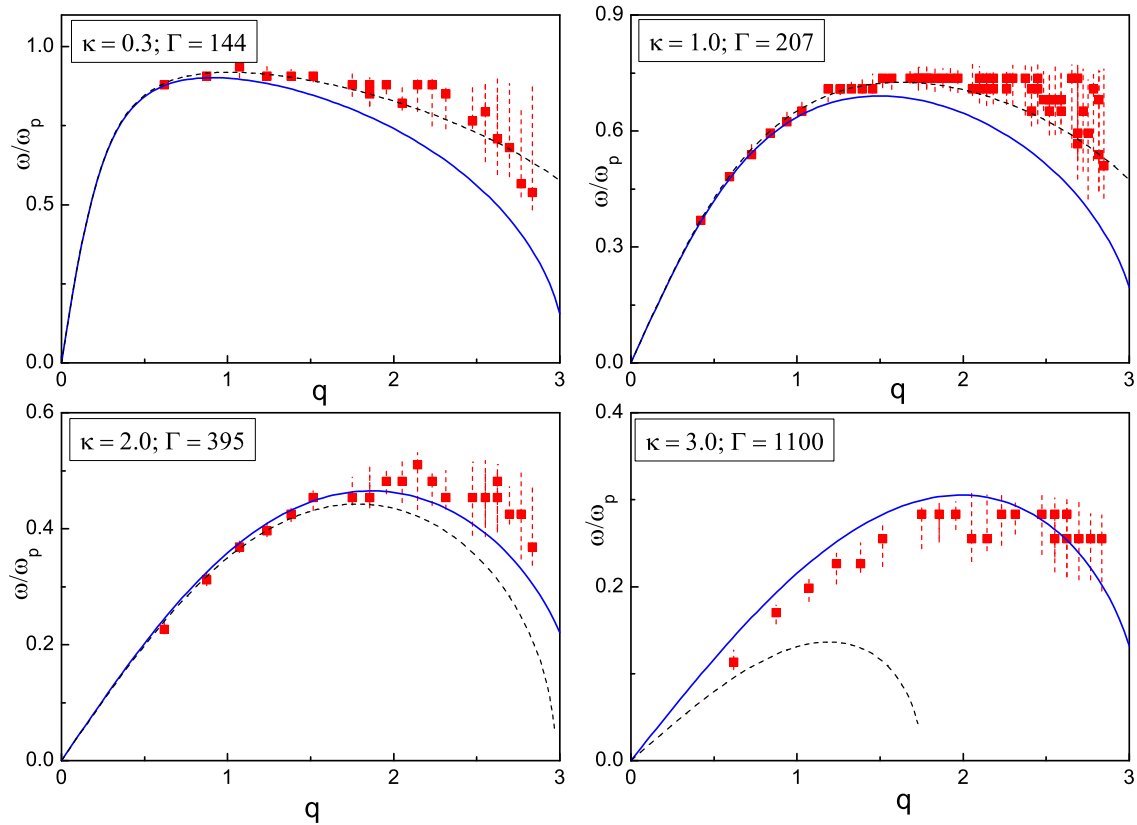}
\caption{(color online) Dispersion of the longitudinal waves in Yukawa fluids near freezing. Symbols correspond to the results from numerical experiment~\cite{MD_DAW}. Solid lines are calculated using Eqs. (\ref{p_ex}) -(\ref{compr}). Dashed curves correspond to the dispersion relation (\ref{OCP_disp}) with the OCP expression (\ref{fit}) for $u_{\rm ex}$. The values of $\kappa$ and $\Gamma$ are given in the upper left corner of each figure. For further details see the text.}
\label{Dispers}
\end{figure*}

Comparison between the numerical results and theory is shown in Fig.~\ref{Dispers}. Symbols and vertical bars represent simulation results for the longitudinal waves and their uncertainties. The solid curves correspond to the dispersion relation (\ref{DR}) with the compressibility evaluated using the DHH approximation (we also assume $\gamma\simeq 1$ for such strong coupling). The dashed curves correspond to the dispersion relation (\ref{OCP_disp}) with the excess energy evaluated from the OCP fit of Eq.~(\ref{fit}).

Figure~\ref{Dispers} demonstrates that the theoretical OCP dispersion relation (\ref{OCP_disp}) fits nicely the numerical data in the regime $\kappa\lesssim 1$. This is clearly the consequence of the fact that the presence of {\it weak} screening does not lead to significant deviations from the equation of state of the OCP, as has been pointed out in Ref.~\cite{Farouki}. However, as screening becomes more pronounced ($\kappa\gtrsim 1$), the dispersion relation (\ref{OCP_disp}) becomes progressively less accurate. At $\kappa=3$ it is completely off the simulation data. Note that this defect cannot be cured by taking into account screening and using more accurate results (either DHH or exact numerical) for the excess energy $u_{\rm ex}$. The actual excess energy {\it decreases} (become more negative) with $\kappa$ (see e.g. Table III from Ref.~\cite{Hamaguchi}), which implies that the situation would become even worse if exact values of $u_{\rm ex}$ are used instead of the OCP results. On the other hand, Eq. (\ref{DR}) is less accurate in the weakly screened regime ($\kappa\lesssim 1$), in particular in the short-wavelength domain, but demonstrates reasonable agreement with the numerical results at stronger screening. Taking into account that the accuracy of the DHH approximation also increases in the strongly screened regime we can conclude that the simplest hydrodynamic description (\ref{DR}) combined with the DHH approximation for the inverse compressibility provides a reasonable compromise between the accuracy and simplicity/speed of calculations in the regime $\kappa\gtrsim 1$.

Another theoretical approach for the waves in strongly coupled Yukawa fluids (dusty plasmas) is based on the quasilocalized charge approximation~\cite{Rosenberg,Kalman}. The theory has been shown to agree very well with the numerical simulation data~\cite{Kalman}. Since this approach also requires evaluation of the system internal energy, simple approximations similar to that considered in the present paper can again be of certain value.

We conclude this section with the following general observation. The excess pressure is a negative decreasing function of $\Gamma$. Thus, the pressure term results in negative contribution to the dispersion relation, provided $\Gamma$ exceeds some critical value. In the OCP limit this transition is known as the onset of negative dispersion and recent numerical simulations locate it at $\Gamma_*\simeq 10$~\cite{Mithen}. [Equations (\ref{DR}) and (\ref{OCP_disp}) yield $\Gamma_*\simeq 5$ and $\Gamma_*\simeq 14$, respectively]. This implies that in the strongly coupled regime the group velocity becomes negative ($\partial\omega/\partial k<0$) at large $k$ as actually seen in all cases shown in Fig.~\ref{Dispers}. This feature is peculiar to the longitudinal modes in solids, which indicates that there is no qualitative difference between the dispersion properties of (strongly coupled) liquid and crystalline Yukawa systems (dusty plasmas). In the long-wavelength limit the longitudinal waves exhibit acoustic behavior ($\omega\propto k$) and their phase velocity is somewhat decreased due to the effect of strong coupling.

\section{Discussion and Conclusion} \label{Concl}

We have discussed a simple analytical approach to estimate the thermodynamic properties of idealized Yukawa systems. The
model considered consists of point-like charges embedded in a neutralizing medium, which is responsible for the exponential
screening and the Yukawa pair interaction potential between the particles. Accurate numerical results exist for this model and these were used for
comparison with our approximation. Although the obtained analytical results do not yield very high accuracy (in particular,
in the regime of strong coupling and weak screening), they provide convenient formulas to describe the essential qualitative
properties of Yukawa systems.

We note, however, that the idealized model does not account for some important properties of real systems. Some of these properties, which
are relevant to complex plasmas are as follows: (i) Particles are not point-like, the typical ratio of the particle size to the plasma screening length can vary in a relatively wide range; (ii) There is a wide region around the particle where the ion-particle interaction is very strong, which results in non-linear screening; (iii)
Particle charge is not fixed, but depends on complex plasmas parameters (e.g. on the particle density); (iv) The average
density of ions and electrons is not fixed, but is related to the particle density and charge via the quasineutrality
condition. Most of these properties are also to a large extent relevant to colloidal dispersions.

Clearly these properties can considerably affect the thermodynamics. From this perspective, accurate results for an idealized
model can be considered as reference data for more advanced models. Extension of simple analytical approximations, similar
to that discussed in this paper, would be a reasonable strategy to study the relative importance of the above mentioned
properties. We leave this for future work.

\begin{acknowledgments}
The authors would like to thank Satoshi Hamaguchi for providing numerical data on wave dispersion relations in Yukawa
fluids. We appreciate funding from the Russian Foundation for Basic Research, Project No. 13-02-01099, and from the European
Research Council under the European Union's Seventh Framework Programme, Grant Agreement No. 267499.
\end{acknowledgments}

\end{document}